\documentclass{article}
\usepackage{amsmath,amssymb}
\usepackage[noblocks]{authblk}
\usepackage{epstopdf}
\usepackage{cite}
\usepackage{siunitx}
\usepackage{color}
\usepackage{ulem}
\usepackage{here}
\usepackage{appendix}
\usepackage{graphicx}% Include figure files
\usepackage{wrapfig}
\usepackage{bm}
\usepackage{comment}
\usepackage{sidecap}
\usepackage[top=28truemm,bottom=28truemm]{geometry}

\begin{document}
\title{
Dynamic behavior of elevators under random inflow of passengers
}
%Author Inf
\author[1,2]{Sakurako Tanida}
\affil[1]{Graduate School of Science, the University of Tokyo, 7-3-1 Hongo, Bunkyo-ku, Tokyo, Japan}
\affil[2]{Research Center for Advanced Science and Technology The University of Tokyo, 4-6-1 Komaba, Meguro-ku, Tokyo, Japan}
%Email
\affil[ ]{\textit {u-tanida@g.ecc.u-tokyo.ac.jp}}
%Date
\date{\today}
\maketitle
%----------------------------------
%%%%%%%%%%%%%%%%%%%%%%%%%%%%%%%%%%%

\begin{abstract}%_______________________________________________________________________________________________________________________

Elevators can be regarded as oscillators driven by the calls of passengers who arrive randomly.
We study the dynamic behavior of elevators during the down peak period numerically and analytically.
We assume that new passengers arrive at each floor according to a Poisson process and call the elevators to go down to the ground floor.
We numerically examine how the round-trip time of a single elevator depends on the inflow rate of passengers at each floor and reproduce it by a self-consistent equation considering the combination of floors where call occurs.
By setting an order parameter, we show that the synchronization of two elevators occurs irrespective of final destination (whether the elevators did or did not go to the top floor).
It indicates that the spontaneous ordering of elevators emerges from the Poisson noise.
We also reproduce the round-trip time of two elevators by a self-consistent equation considering the interaction through the existence of passengers and the absence of volume exclusion.
Those results suggest that such interaction stabilizes and characterizes the spontaneous ordering of elevators.
\end{abstract}

\section{Introduction}
Congestion is one of the challenging problems we face in everyday life.
Physically it can be described as a process in a non-equilibrium system~\cite{Nagatani2002a,Helbing2001,Chowdhury2000b}. 
or example, the traffic jam was empirically known to emerge even in normal conditions, without disruptions caused by traffic accidents or bottlenecks.
The quantitative measurements revealed that the jamming transition occurred as the vehicle density increased~\cite{Hall1986,Neubert1999}.
More specifically, the traffic flow increased linearly with the density at low densities, while widely varying at high densities.
Interestingly, for the intermediate densities, the flow did not uniquely depend on the density.
The latter implied the existence of the hysteresis effect and meta-stable states.
The traffic jam was also known for showing density waves and several models were proposed to describe this phenomenon~\cite{Musha1976,Musha1978,Kerner1993,Komatsu1995,Kurtze1995}.
Generally speaking, out-of-equilibrium behaviors are often observed in the transportation systems, where traffic participants move and interact with each other.

Elevators represent one more example of transportation systems exhibiting nontrivial out-of-equilibrium behaviors.
Multiple elevators form clusters in the crowded system and thus its density separation is similar to traffic jams~\cite{Poschel1994,Hikihara1997,Nagatani2002,Nagatani2003,Nagatani2004,Nagatani2011,Nagatani2012,Nagatani2015,Nagatani2016,Feng2020}.
However, different from traffic jams, elevators exhibit spontaneous ordering as the out-of-equilibrium oscillators.
Elevator and car systems are also different in a way they interact with other traffic participants.
More specifically, elevators freely pass other elevators, but cars cannot go ahead if other cars are present in front of them.
Such a difference originates from whether vehicles have the volume exclusion effect.
Another difference appears in the control parameter.
While the traffic jam emerges when the car density increases, the clustering of elevators originates from the increasing mean number of passengers coming in a unit time.
The dynamics of elevators’ motion as a function of inflow was investigated for various scenarios.
More specifically, the pioneering numerical studies reported the synchronization of multiple elevators moving down during peak loads~\cite{Poschel1994, Hikihara1997}.
The interaction between the elevators, however, was not clarified, and the mechanisms of the synchronization remained elusive.
Recently, stochastic models have been developed and applied to estimate the typical time of round trips in a steady state for each inflow~\cite{Nagatani2003,Nagatani2004,Nagatani2011,Nagatani2012,Nagatani2015,Nagatani2016}.
These models assumed that elevators’ motion presented a limit cycle with the necessary stops at both top and ground floors.
However, the motion of the elevator system is essentially caused by the Poisson noise of passengers’ arrivals.
Thus, in order to understand the mechanism of elevator clustering, it is important to investigate its spontaneous ordering that emerges from this noise.

In this paper, we investigate elevators’ downward motion during peak loads when passengers go down to the ground floor to exit the building.
We numerically simulate the motion of a single elevator in a building where the passengers arrive according to the Poisson process.
We measure the round-trip time for various inflow rates of passengers and reproduce it by an equation considering the combination of floors where call occurs. 
Next, we examine the simultaneous arrivals of two elevators by introducing an order parameter.
We show that the simultaneous arrivals of elevators occur irrespective of final destination (whether the elevators do or do not go to the top floor).
Moreover, we demonstrate that the round-trip time of two elevators can be estimated by a self-consistence equation considering the interaction between elevators and the absence of the volume exclusion effect.
Finally, we show that the interaction stabilizes the spontaneous ordering of elevators, and the absence of the volume exclusion characterizes the dynamics of the cluster.
Note that elevator system efficiency considerations are beyond the scope of this study.

%_______________________________________________________________________________________________________________________
\section{Problem formulation}
Let us consider a downward elevator system motion during peak loads.
We assume that the arrival of new passengers at each floor is a Poisson process.
Moreover, the passenger inflow rate at each floor is uniform, and elevators do not necessarily serve the top floor.
To model the dependence of a round-trip time from the inflow rate, we use a self-consistent equation considering combinations of calling floors during a round-trip.

Our elevator system consists of one or two elevators serving $K$ floors (Fig.~\ref{fig:schematic}).
The elevators take 1 time step to move 1 floor up or down and $\gamma$ time steps for the passengers to enter or exit.
In this paper, we set $\gamma=10$ for all cases.
Consistent with previous studies \cite{Poschel1994,Hikihara1997}, we assume that all calls for elevators are from passengers waiting for the elevators at floor $1\sim K$ in order to move to the ground floor and exit the building.
Once an elevator goes down, it does not go up again until it arrives at the ground floor. 
If there are no passengers in neither elevators nor floors, the elevators stay at the floor where they stopped until the next call. 
In case of no waiting calls and no passengers in the elevators, the next call will be accepted by the elevator closest to the calling floor.
Provided there are multiple (more than one) unresolved calls and both elevators are free, the elevator stopped at a higher floor (the upper elevator) moves upward to the higher calling floor. 
Simultaneously, the lower elevator starts moving if there are some calling it can reach faster compared to the upper one. 
Each elevator can simultaneously carry not more than $M$ passengers.
The elevators are not smart enough to identify the number of carrying passengers and possibly stop even if they are full.
Unless otherwise noted, we focus on the case that the capacity is large enough not to be filled up, $M=10000$. Finally, we set the initial position of elevators randomly and performed each simulation for at least 10000 time steps.

% FIG %%%%%%%%%%%%%%%%%%%%%%%%%%%%%%%%%%%%%%%%%%%%
\begin{figure}[H]%{r}[0pt]{0.5\textwidth}
\centering
\includegraphics[width=.5\linewidth]{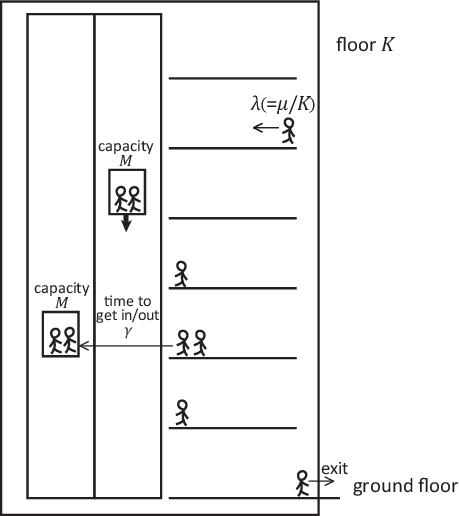}
\caption{Schematic illustration of this model.
}
\label{fig:single}
\end{figure}
%%%%%%%%%%%%%%%%%%%%%%%%%%%%%%%%%%%%%%%%%%%%%

To estimate the time to go to the target floor, $k_t$, from the floor, $k$, we make the following additional assumptions: 
1)	If the elevator is going up or at the ground floor, it will stop at the highest floor, $k_h$, among those having unresolved calls and then will switch to going down and will stop at all floors with unresolved calls between $k_h$ and $k_t$.
2)	If the elevator is going down and $k>k_t$, it will stop at all the floors with unsolved calls between $k$ and $k_t$.
3)	If the elevator is going down and if $k_t \leq k$, it will stop at all the floors with unresolved calls between $k$ and ground floor and between $k_h$ and $k_t$.
The estimated time is also used to decide whether the lower elevator goes up or stops at a floor with unresolved calls, in case of the upper elevator having passengers and going down.
Note that the actual time to arrive at the target floor is different from the estimated one because the passenger inflow is updated at every time step.
The number of new passengers at each floor and at every time step $n$ is distributed according to Poisson law :
\begin{eqnarray}
	P_{\lambda}(n) = \frac{\lambda^n}{n!}e^{-\lambda} \ ,
	\label{eq:poisson}
\end{eqnarray}
where $\lambda=\mu/K$ is the Poisson parameter and $\mu$ stands for the passenger inflow rate for the entire building.

\clearpage
%_____________________________________________________________________________________________________________
\section{Results}
First, we examined the relation between the passenger inflow rate, $\mu$, and the behavior of an isolated single elevator in a 10-floor building ($K=10$).
Figures \ref{fig:single}(a)-\ref{fig:single}(d) show examples of the time evolution of the elevator's position.
In general, if $\mu$ is small, the elevator stays at the ground floor and resolves only one call for most round trips.
As $\mu$ increases, the elevator makes round trips between the ground floor and the upper floor continuously.
Fig.~\ref{fig:single}(e) shows the percentage of the waiting time at the ground floor $r$ for various values of $\mu$.
As $\mu$ increases, the value of $r$ asymptotically approaches $\gamma/130$.
Note that 130 is the time of a round trip when the elevator stops at all floors including the ground floor.
More specifically, it decays exponentially as $\exp(-\mu/0.03)$ for small $\mu$ and switches to $\exp(-\mu/0.07)$ for $\mu\gtrsim 0.1$.

% FIG %%%%%%%%%%%%%%%%%%%%%%%%%%%%%%%%%%%%%%%%%%%%
\begin{figure}[H]
\centering
\includegraphics[width=.5\linewidth]{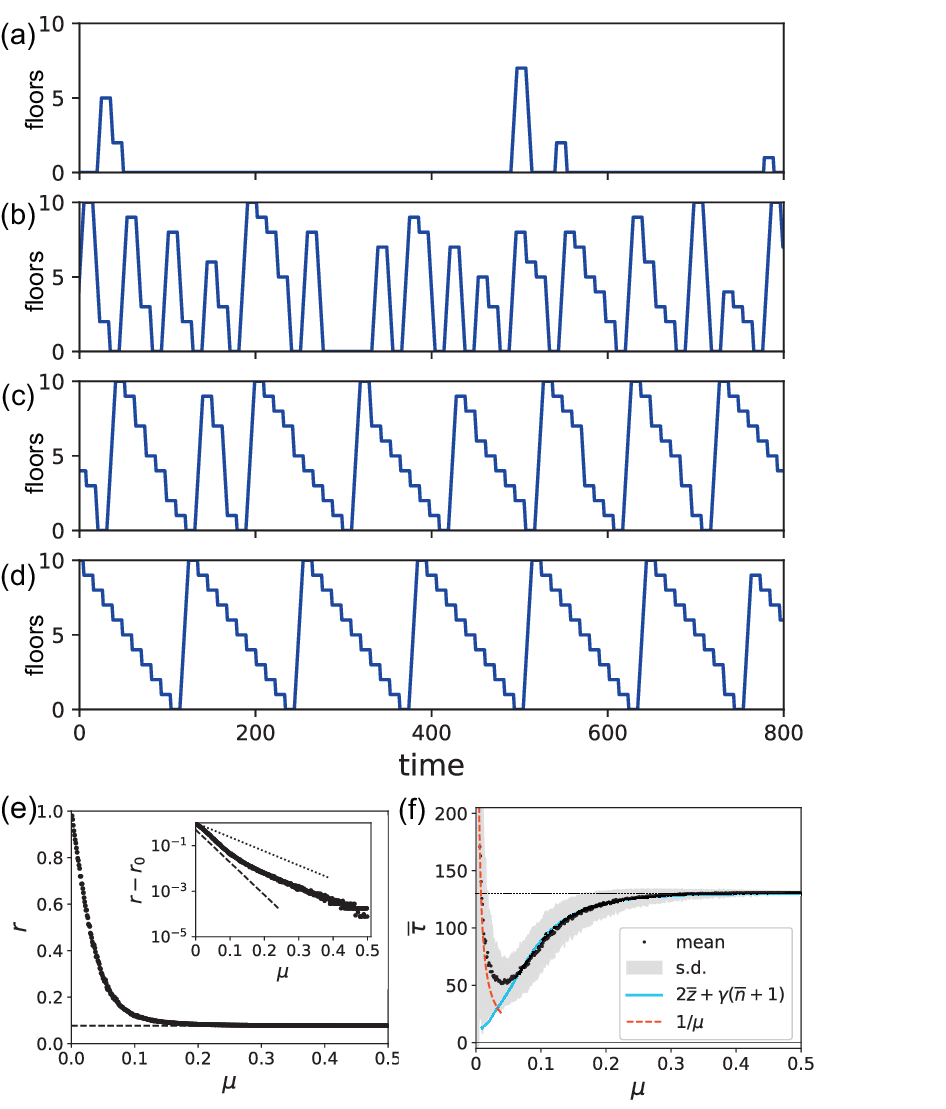}
\caption{[(a)-(d)] Single elevator dynamics for small passenger inflow rates: (a) $\mu=0.01$, (b) $\mu=0.04$, (c) $\mu=0.06$, and (d) $\mu=0.2$. The dark blue lines represent the position of the elevator at each time step. As $\mu$ increases, the elevator moves periodically.
(e)~Percentage of the elevator waiting time at the ground floor, $r$, vs the passenger inflow rate, $\mu$, for $K=10$. As $\mu$ increases, $r$ approaches the asymptotic value of $r_0=\gamma/130$ (broken line). Inset: $r-r_0$ decays exponentially as $\mu$ increases. The broken and dot lines represent slopes of $\exp(-\mu/0.03)$ and $\exp(-\mu/0.07)$, respectively.
(f)~The mean time (black dots) between two consecutive departures from the ground floor and corresponding standard deviation (shaded area). The mean initially decreases for small $\mu$ and starts increasing at about $\mu=0.04$. The solid cyan and broken orange lines represent estimated values obtained from Eq.(\ref{eq:timetoestimate}) and an exponential distribution, respectively. The dotted black line represents $\overline{\tau}=130$.
}
\label{fig:single}
\end{figure}
%%%%%%%%%%%%%%%%%%%%%%%%%%%%%%%%%%%%%%%%%%%%%

Figure~\ref{fig:single}(f) shows the mean time between two consecutive departures from the ground floor, $\overline{\tau}$.
For small $\mu$, $\overline{\tau}$ decays as $1/\mu$ because the elevator moves to resolve a single call.
Its interval follows an exponential distribution [orange broken line in Fig.\ref{fig:single}(f)].
The mean time, $\overline{\tau}$ reaches the minimum for $\mu=0.04$ and starts increasing for higher values of passenger inflow rates.
The maximum value of the mean time is $\overline{\tau}=130$ (time steps) and corresponds to the elevator the starts from the ground floor and stops at all floors.
Turning into an increase from the $1/\mu$ decay implies that the elevator motion shifts from the intermittent to the periodic motion.

Let us now define $\overline{\tau}$ for the periodically moving elevator.
The total time of the $j$-th round trip, $\tau_j$ (time interval between the $j$-th and the $(j+1)$-th departures) reads as $\tau_j = 2 z_j + \gamma( n_j+1)$. Here $z_j$ is the highest floor in the $j$-th round trip, and $n_j$ is the number of floors where the elevator stops.
Note that adding 1 in the second term corresponds to a stop at the ground floor.
Let $m_j$ be the number of new passengers coming during the $j$-th round trip.
Then the number of new passengers coming during N($\gg 1$) round trips can be approximated as $\sum_{j=0}^N m_j = \mu \sum_{j=0}^N \tau_j$.
Thus, the mean number of new passengers on a round trip is:
\begin{eqnarray}
	\overline{m} = \mu [2\overline{z}+\gamma(\overline{n}+1)]\ ,
    \label{eq:timetoestimate}
\end{eqnarray}
where $\overline{x}=1/N\sum_{j=0}^N x_j$.

\begin{figure}[H]
\centering
\includegraphics[width=.5\linewidth]{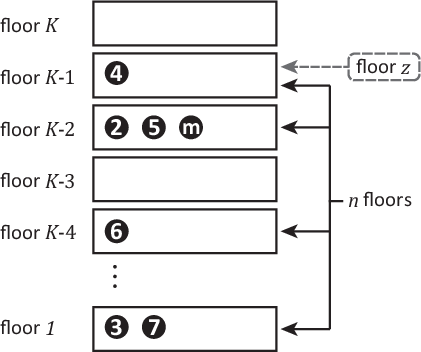}
\caption{Schematic illustration of the ball-in-box problem corresponding to our model when $m$ passengers are waiting for elevators in a $K$-floors building.
The passengers are labeled according to the order of arrival.
Black circles represent individual passengers.
The highest floor where passengers are waiting is $z$ and the number of floors where passengers are waiting is $n$.
}
\label{fig:combi}
\end{figure}

The value of $\overline{z}$ is estimated by $E_{\overline{m}}(z)$, the expectation of the highest floor where the passengers are waiting provided that $\overline{m}$ passengers are randomly located at $K$ floors.
This expectation is following the methodology of the ball-in-box problem (Fig.\ref{fig:combi}). 
The number of ways $m$ passengers can be located at $k$ floors is $k^{\overline{m}}$. Then combination satisfying $k$ being the highest floor is $k^{\overline{m}}-(k-1)^{\overline{m}}$.
Thus, the expectation of the highest floor among those stopped during a round trip is
\begin{eqnarray}
    E_{\overline{m}}(z) %\overline{z} 
    &=& \frac{1}{ K^{\overline{m}} }\sum_{k=1}^K k[ k^{\overline{m}}-(k-1)^{\overline{m}} ] \nonumber \\
    &=& K-\frac{1}{K^{\overline{m}}}\sum^{K-1}_{k=1} k^{\overline{m}} \ .
     \label{eq:z}
\end{eqnarray}

% FIG %%%%%%%%%%%%%%%%%%%%%%%%%%%%%%%%%%%%%%%%%%%%
%%%%%%%%%%%%%%%%%%%%%%%%%%%%%%%%%%%%%%%%%%%%%

Here $\overline{n}$ can be considered as the expectation value of the number of floors where more than zero passengers arrive during a round trip.
Assuming that the typical round-trip time is $\tau$ and $\overline{m}=\mu\overline{\tau}$, the expectation value ,$E_{\overline{m}}(n)$ reads as:
\begin{eqnarray}
    E_{\overline{m}}(n) &=& K\left(1-e^{-\overline{m}/K} \right)
    \ .  \label{eq:n}
\end{eqnarray}

Because both Eqs.(\ref{eq:z}) and (\ref{eq:n}) are the functions of $m$, Eq.(\ref{eq:timetoestimate}) is the closed form of $\overline{m}$.
As shown in Fig.~\ref{fig:single}(h) (solid line), the value of $2\overline{z}+\gamma (\overline{n}+1)$ derived from Eq.(\ref{eq:timetoestimate}) for each $\mu$ can appropriately reproduce the simulation results for $\mu$ greater than 0.05.
It indicates that the elevator motion is periodic for this range of the passenger inflow rate.
Additionally, to satisfy the ``no new calls during the round trip" condition, $\mu<0.03$ should be satisfied that corresponds to $n<1$.
For $\mu$ ranging from 0.03 to 0.05, the intermediate motion emerges.

% FIG %%%%%%%%%%%%%%%%%%%%%%%%%%%%%%%%%%%%%%%%%%%%
\begin{figure*}[htb]
\centering
\includegraphics[width=.99\linewidth]{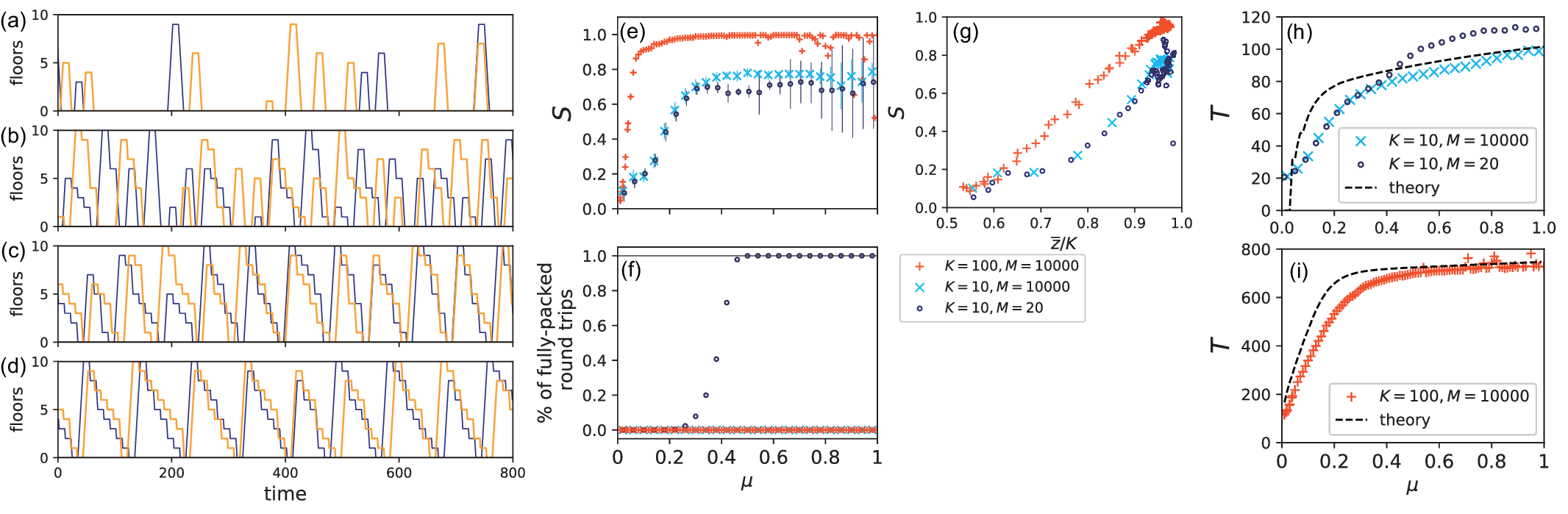}
\caption{[(a)-(d)] Two elevators dynamics for different values of the passenger inflow rate: (a) $\mu=0.02$, (b) $\mu=0.1$, (c) $\mu=0.2$, and (d) $\mu=0.4$. Dark blue and bright orange lines represent the positions of each elevator at a given time step. As $\mu$ increases, the elevators move as a cluster.
(e)~The simultaneousness ($S$) as a function of the passenger inflow rate. Cross and blue dot markers represent mean $S$ for$(K,M)=(10,10000)$ and $(10,20)$, respectively. The vertical lines on the cross and dot markers represent the standard deviation. The red plus markers represent mean $S$ for $(K,M)=(100,10000)$.
(f)~Percentage of fully-packed round trips as a function of the passenger inflow rate. Each marker style is the same as those in~(e). Although $S$ shows the similar behavior for $(K,M)=(10,20)$ and $(10,10000)$, only the percentage of $(K,M)=(10,20)$ rises with $\mu$.
(g)~The simultaneousness as a function of the normalized highest floor in a round trip ($\overline{z}/K$). Here $\overline{z}$ is the mean of the highest floors where the elevators stop in a round trip, and $K$ is the highest floor of the building.
[(h)-(i)] ~ Time interval between two consecutive elevators arrivals at the ground floor ($T$) as a function of the passenger inflow rate. Numerical and theoretical plots for $K=10$ and $K=100$ are given in~(g) and~(h), respectively. Each marker style is the same as those in~(e). The broken lines represent theoretical values derived from Eq.(\ref{eq:timetoestimate2}).}
\label{fig:double}
\end{figure*}
%%%%%%%%%%%%%%%%%%%%%%%%%%%%%%%%%%%%%%%%%%%%%

Next, the dynamics of the two elevators was investigated.
It is shown in Figs.~\ref{fig:double}(a-d) for $\mu = 0.02, 0.1, 0.2,$ and 0.4.
Similar to the case of a single elevator, the elevators stay at the ground floor when $\mu$ is small [Fig.~\ref{fig:double}(a)].
The elevators exhibit round trips between the ground floor and the upper floor continuously when $\mu$ is in intermediate range [Fig.~\ref{fig:double}(b)].
When $\mu$ is large, the two elevators arrive at the ground floor simultaneously as $\mu$ increases [Figs.~\ref{fig:double}(c-d)].
To quantify this tendency, we measure the simultaneousness, $S$, defined as follows: 
\begin{eqnarray}
	S = \left| \frac{1}{N}\sum^{N}_{j=1} \exp \left(2\pi i \frac{ \tau_{j}}{\overline{\omega}}\right)\right| \ ,
\end{eqnarray}
where $\tau_j$ is the time between $j$-th and ($j+1$)-th departures of either elevator at the ground floor, $\overline{\omega}$ is a mean time of a round trip from the departure to the next arrival of both elevators, and $N$ is a total number of departures.
The dot and cross markers in Fig.~\ref{fig:double}(e) show mean $S$ for parameter sets of $(K,M)=(10,20)$ and (10,10000), respectively.
In both cases, $S$ rapidly increases for small $\mu$.
Above $\mu=0.3$, mean S of $(K,M)=(10,20)$ is fluctuated around $S=0.7$.
The vertical lines on dot and cross makers show the standard deviation of S for each $\mu$.
The standard deviation of $(K,M)=(10,20)$ is large above $\mu=0.6$.
As shown in Fig.~\ref{fig:double}(f), for $(K,M)=(10,20)$, the percentage of fully-packed round trips is almost 0 below $\mu=0.25$, increases above 0.25, and reaches 1 at $\mu=0.4$.
Those results suggest that, for a high inflow rate, the queues of the waiting passengers remain, and the dynamics of elevators becomes deterministic.
In such cases, the initial condition affects the degree of simaltaneous arrival.
On the other hand, for $(K,M)=(10,10000)$, mean $S$ reaches saturation above $\mu=0.3$, the standard deviation is small until $\mu=0.8$, and the percentage of fully-packed round trips stays 0 for all $\mu$.
These results indicate that the simultaneous arrival can occur even in the case that the elevators are not filled.

We also investigate behavior of mean $S$ at $(K,M)=(100,10000)$ and find that it rises for smaller values of $\mu$ than and stabilize from $\mu=0.2$ [red plus markers in Fig.~\ref{fig:double}(e)].
The values of $S$ fluctuate for $\mu$ exceeding 0.5, implying that the initial distance between two elevators does not get smaller.
It can be due to the dynamics of elevators becomes close to deterministic.
The fully-packed percentage remains 0 when $\mu \leq 1$ [Fig.~\ref{fig:double}(f)].

To figure out whether the elevators arrive simultaneously even if they do not reach the top floor, we examine the relation between $\overline{z}/K$ and $S$ for various values of $\mu$.
As shown in Fig.~\ref{fig:double}(g), $S$ rises when $z$ is relatively small.
The simultaneous arrival occurs without elevators' reaching the top floor.
It indicates that the simultaneous arrival can emerge from the noise even without the given limit cycle.

We define $T$ as the round-trip time from an elevator's departure from the ground floor to the elevator's next arrival on the ground floor.
As the time since the last elevator departure is shorter, fewer new passengers come at a given floor.
Hence, the number of floors where the elevator stops becomes smaller, its speed increases and finally overcomes that of the leading elevator.
As a result, the distance between the two elevators reduces with time.
Because the two elevators are symmetric and the elevator system has no volume exclusion problem, the anteroposterior relation of the two elevators is frequently switched when the distance is close to 0.
Therefore, it can be assumed that about half of the elevator calls occurred during the last round trip are resolved by an elevator, and the other half is resolved by another elevator.
Additionally, the probability of new passengers coming at the floor since the forward elevator departed from that floor is assumed to be $1-\exp(-\gamma\mu/K)$. The latter can be attributed to the fact that the typical time since the forward elevator has left is $\gamma$.
Hence, assuming the two elevators move as a cluster, the number of floors where the elevator stops during a round trip can be represented as $\frac{n}{2}+ \left(z-\frac{n}{2}\right)
\left(1-e^{ -\frac{\gamma\mu}{K}}\right)$, where $z$ is the highest calling floor for the last round trip of the cluster, and $n$ is the number of calling floors for the last round trip of the cluster.
Similar to derivation of Eq.(\ref{eq:timetoestimate}), the mean number of new passengers in a round trip is given by:
\begin{eqnarray}
	\overline{m} = \mu \left\{ 2\overline{z}+\gamma\left[
	\frac{\overline{n}}{2}+1+
	\left(\overline{z}-\frac{\overline{n}}{2}\right)
	\left(1-e^{ -\frac{\gamma\mu}{K}}\right) 
	\right]\right\} \ .
    \label{eq:timetoestimate2}
\end{eqnarray}
As $\overline{z}$ and $\overline{n}$ are estimated as the functions of $\overline{m}$ [see Eqs.(\ref{eq:z}) and (\ref{eq:n})], Eq.(\ref{eq:timetoestimate2}) also presents closed form of $\overline{m}$.
As shown in Fig.~\ref{fig:double}(h), $T$ calculated from Eq.(\ref{eq:timetoestimate2}) reproduces the trend obtained in simulations.
As the assumption of the distance between two elevators being always small was inadequate for small $S$ values, the difference between theory and simulation was relatively large for $\mu$ less than 0.2.
The value of $T$ of $(K,M)=(10,20)$ above $\mu=0.4$ mismatch that of obtained from Eq.(\ref{eq:timetoestimate2}).
It suggests that the theory can be applied when the elevators are not fully-packed.

%_____________________________________________________________________________________________________________
\section{Discussion and Conclusions}
We have examined the dynamics of elevators moving downwards for passenger inflow corresponding to a Poisson process.
We defined the time between two consecutive departures from the ground floor of a single elevator for each level of the passenger inflow rate and found that the relation of those could be separated into two.
For a small inflow rate, the time decayed as inflow increased and it followed the mean of the exponential distribution.
For a large inflow rate, the time increased with inflow and was reproduced by the self-consistent equation considering possible combinations of floors on which the calling occurs during a round-trip time.
Next, we numerically studied the conditions for the simultaneous arrival of two elevators by introducing the order parameter.
We showed that the simultaneous arrival can occur even in the case of the unlimited capacity.
We also found that the elevators would arrive simultaneously even if the elevators did not go to the top floor. 
It indicated that the spontaneous ordering of elevators emerged from the Poisson noise.
We defined the round-trip time for each level of the passenger inflow rate and reproduced the round-trip time by a self-consistent equation based on the equation for a single elevator.
For the self-consistent equation of two elevators, we assumed that the smaller the distance between the two elevators is the faster the posterior elevator becomes.
Such interaction where the distance between vehicles is small is the same as that of buses in the bus-route model~\cite{OLoan1998,Nagatani2001,Gershenson2009}.
However, as mentioned in the Introduction, the volume exclusion effect does not work for elevators, while it does for the bus-route model.
Hence, the elevators could overcome others and the anteroposterior relation of the two elevators was frequently switched when the distance was close to 0.
It results in the mean speed of multi elevators can be faster than that in a single elevator system, while that of buses cannot be.
Thus, we can conclude that the interaction stabilized the spontaneous ordering of elevators, and the absence of the volume exclusion characterized the dynamics of the cluster.
Those results clarify the mechanisms of elevator clustering and will open new venues for optimization of passenger flows in tall buildings.

\section*{Acknowledgment}
We are grateful to Tetsuya Hiraiwa, Hiroshi Kori, Katsuhiro Nishinari, Claudio Feliciani, Daichi Yanagisawa, and Hisashi Murakami for helpful discussions and for their kind interest in this work.

%\bibliographystyle{unsrt}
%\bibliography{library}

\end{document}